\documentclass[iop]{emulateapj}

\def\viibf{}
\def\mybf{}
\def\newbf{}
\def\newtext{}

\def\halpha{H$\alpha$}

\def\sqig{$\sim$}

\def\ctscm2s{cts\,cm$^{-2}$\,s$^{-1}$}

\def\Chandra{{\it Chandra}}
\def\Fermi{{\it Fermi}}
\def\INTEGRAL{{\it INTEGRAL}}
\def\RXTE{{\it RXTE}}
\def\Swift{{\it Swift}}
\def\XMM{{\it XMM-Newton}}

\def\cts{counts~s$^{-1}$}
\def\halpha{H$\alpha$}

\def\Porb{P$_{orb}$}
\def\Ps{P$_{pulse}$}

\def\igr1448{IGR\,J14488-5942}
\def\igrj1448{IGR\,J14488-5942}
\def\ax1700{AX\,J1700.2-4220}
\def\axj1700{AX\,J1700.2-4220}
\def\swiftj1816{Swift\,J1816.7-1613}

\begin{document}

\def\subtitle{}
\submitted{}
\accepted{August 10, 2017}
\journalinfo{}

\title{
Diverse Long-Term Variability of Five Candidate High-Mass X-ray Binaries
from \Swift\ Burst Alert Telescope Observations}

\author{Robin H.~D. Corbet\altaffilmark{1,2,3}}
\author{Joel B. Coley\altaffilmark{4,5}}
\author{Hans A. Krimm\altaffilmark{6,7}}

\altaffiltext{1}{University of Maryland, Baltimore
County, MD 21250, USA; corbet@umbc.edu}

\altaffiltext{2}
{CRESST/Mail Code 662, X-ray Astrophysics Laboratory,
NASA Goddard Space Flight Center, Greenbelt, MD 20771, USA}

\altaffiltext{3}
{Maryland Institute College of Art, 1300 W Mt Royal Ave, Baltimore, MD 21217, USA}

\altaffiltext{4}
{NASA Postdoctoral Program, and Astroparticle Physics Laboratory,
Code 661 NASA Goddard Space Flight Center, Greenbelt Rd., MD 20771, USA.}

\altaffiltext{5}
{CRESST/Astroparticle Physics Laboratory, 
Code 661 NASA Goddard Space Flight Center, Greenbelt Rd., MD 20771, USA}

\altaffiltext{6}
{Universities Space Research Association,
10211 Wincopin Circle, Suite 500, Columbia, MD 21044, USA}

\altaffiltext{7}
{National Science Foundation}

\begin{abstract}
We present an investigation of
long-term modulation in the X-ray light curves of five little-studied
candidate high-mass X-ray binaries using the \Swift\ Burst Alert Telescope.
\igrj1448\ and
AX J1700.2-4220 show strong modulation at periods of 49.6 and 44 days, respectively, which
are interpreted as orbital periods of Be star systems.
For \igrj1448, observations with \Swift\ X-ray Telescope
show a hint of pulsations at 33.4 s.
For AX J1700.2-4220, 54 s pulsations were previously 
found with \XMM. 
Swift J1816.7-1613 exhibits complicated behavior. The strongest peak
in the power spectrum is at a period near 150 days, but this conflicts
with a determination of a period of 118.5 days by La Parola et al. (2014).
AX J1820.5-1434 has been proposed to exhibit modulation near 54 days, but
the extended BAT observations suggest modulation at slightly
longer than double this at approximately 111 days. There appears to be
a long-term change in the shape of the modulation near 111 days, which
may explain the apparent discrepancy.
The X-ray pulsar XTE J1906+090, which was previously proposed to be a Be star system
with an orbital period of \sqig 30 days from pulse timing, shows peaks in the power spectrum at 81 and 173
days. The origins of these periods are unclear, although they might be the
orbital period and a superorbital period respectively.
For all five sources, the long-term variability, together with the combination of orbital
and proposed pulse periods, suggests that the sources contain Be star mass donors.

\end{abstract}

\section{Introduction}

High-mass X-ray Binaries (HMXBs) consist of a neutron star or black hole
accreting from an early spectral type (O or B) companion.
Mass transfer can occur in a variety of ways, depending on the nature of
the mass donor and the orbital separation of the components \citep[e.g.][]{Walter2015}. 
HMXBs show variability on various timescales, these include
the rotation (``pulse'') period of the accreting neutron star and the
orbital period. 

The largest sub-group of
HMXBs is those where accretion occurs from a Be star \citep[e.g.][]{Charles2006,Reig2011}. A Be star is
a non-supergiant OB star that has, at some time, displayed emission
at H$\alpha$ \citep[e.g.][]{Porter2003}. These stars posses a circumstellar ``decretion'' disk,
which is believed to be related to the rapid, but sub-critical, rotation of the star,
{\mybf and possibly also non-radial stellar pulsations \citep[e.g.][and references therein]{Rivinius2013}.}
{\mybf It is from this disk, when it is present, that
accretion may occur.} The circumstellar disk is not persistent,
and may appear and dissipate on timescales of years.
In other HMXBs, accretion takes place
from an OB supergiant component and accretion may be either from the stellar
wind or, in a small number of systems, via Roche-lobe overflow \citep[e.g.][]{Chaty2010}.

In Be star systems, outbursts can be seen which are generally classified
into two categories \citep[e.g.][]{Kretschmar2012}. ``Type I'' outbursts recur with the orbital
period and are expected to occur near periastron passage.
``Type II'' outbursts, which are irregular and occur less frequently, are more luminous, are not obviously related to the
orbital phase, and can have longer durations.
The cause of Type II outbursts is unclear, but suspected to be related
to the properties of the Be star decretion disk, which is thought to be
tidally truncated at a radius resonant with the neutron star orbit \citep[e.g.][]{Reig2016}. 
\citet{Monageng2017} searched for a correlation between decretion disk size and
Type II outbursts from five Be star HMXBs and did not find this. Instead, \citet{Monageng2017}
considered other changes in disk properties, precession, elongation and density effects,
that could account for Type II outbursts.

It is also possible for a Be star HMXB to show apparently regular
outbursts for a time which do not recur on the orbital period.
For example, a \sqig 80 day period was found in the X-ray flux 
from XTE\,J1946+274 by \citet{Campana1999}, while pulse timing subsequently
showed the orbital period to be \sqig 170 days, with the difference
in periods perhaps related to an offset between the orbital plane
and the plane of the Be star's decretion disk \citep{Wilson2003,Marcu2015}.

In Be star systems, long-timescale variations associated with
changes in the circumstellar disk may be seen.
In all sub-groups of HMXBs, superorbital periods can be seen in some, but not all, 
systems \citep[e.g.][and references therein]{Raj2011,Corbet2013}.
For Be star systems, at least some of this long term variability may be attributed to
one-armed oscillations in the decretion disk \citep[e.g.][]{Okazaki1991,Ogilvie2008,Okazaki2016}.

Orbital and long-timescale changes in HMXBs are well suited to study with
X-ray all-sky monitors. Such an instrument is the Burst Alert Telescope (BAT) on board
the \Swift\ satellite \citep{Gehrels2004}, which has now been operating for
approximately 12 years. The hard ($>$ 15 keV) X-ray light curves produced by the BAT
are particularly important for HMXBs located on the Galactic plane{\mybf ,} which can
suffer from significant amounts of interstellar and/or local absorption.
Light curves from the BAT have enabled the discovery of a number of orbital and super-orbital 
periods for HMXBs. In our previous work on long-term BAT data
we have presented analyses of {\viibf symbiotic X-ray binaries \citep{Corbet2008}}, superorbital modulation in supergiant HMXBs
\citep{Corbet2013}, and measurements of eclipses in HMXBs \citep{Coley2015}.
{\newbf We routinely monitor BAT light curves, and the power spectra of these,
for indications of periodic modulation, particularly for sources located on the Galactic plane.}
{\newbf Based on this monitoring,} in this paper we select five poorly-studied candidate Be-star HMXBs for which the
{\newbf BAT power spectra} show features which {\newbf suggest} either orbital,
or other types of long-term, modulation.
These sources are: \igr1448, \ax1700, Swift J1816.7-1613, AX J1820.5-1434, 
and XTE J1906+090. These sources display a variety of behavior ranging from
apparently clear orbital modulations to more complicated variability.

For none of the sources are there light curves from MAXI \citep{Matsuoka2009}, 
or \Fermi\ Gamma-ray Burst Monitor occultation observations
\citep{Wilson2012}, and so the BAT provides the only long-term hard X-ray light curves.
For \ax1700\ a long duration X-ray light curve is also available from {\it Rossi X-ray Timing
Explorer} (\RXTE) Proportional Counter Array (PCA) observations.
Unless otherwise stated, all uncertainties are given at the 1$\sigma$ level.
Preliminary reports of the orbital periods in \igr1448\ and \ax1700\ appeared in
\citet{Corbet2010b}, and \citet{Corbet2010a} respectively, but the results presented here are based on light curves
covering almost twice the durations of those in the early reports.

\section{Data and Analysis}

\subsection{Swift BAT Light Curves}

The \Swift\ BAT is
described in detail by \citet{Barthelmy2005},
and descriptions of our previous use of data from the BAT can 
be found in \citet{Corbet2013} and \citet{Coley2015}.
The BAT uses a coded mask to provide a wide field of view (1.4 sr half-coded, 2.85 sr 0\% coded),
which facilitates its use as an all-sky monitor. 
The pointing direction of \Swift\ is driven by
the narrow-field XRT and UVOT instruments on board \Swift{\mybf ,} resulting
in the BAT typically observing 50\%--80\% of the sky each day.
We use data from the \Swift\ BAT transient monitor \citep{Krimm2013}, which
are available shortly after observations have been performed and
cover the energy range 15 - 50 keV. 
In this energy range, the Crab gives a count rate of 0.22 \ctscm2s.
Transient monitor light curves
are available with time resolutions of \Swift\ pointing durations (``orbital light curves''), and also daily averages.
BAT transient monitor light curves are currently available for 1281 sources.
The light curves of the five sources 
considered here cover the time range of MJD 53,416 to 57,673 (2005-02-15 to 2016-10-12).
While BAT light curves are also available from catalogs \citep[e.g.][]{Tueller2010,Baumgartner2013},
the transient monitor light curves are of much longer duration. 
As in our previous analyses \citep{Corbet2013,Coley2015},
we used only data for which the data quality flag (``DATA\_FLAG'') was
0, indicating good quality and, in addition,
removed points with very low fluxes and implausibly small uncertainties from the light curves.
Because the BAT pointing direction is primarily driven by XRT and UVOT observations, the observation
durations and frequency of durations is variable.
The individual observations in the orbital light curves of the fives sources have exposures ranging
from 64 to 2664 s, with mean exposures of \sqig630 s.

\subsection{Period Searches using Power Spectra}

Our searches for periodicity in the BAT light curves
also followed the procedures described in \citet{Corbet2013}.
Our primary method for searching for periodic modulation is to use
discrete Fourier transforms (DFTs) to obtain power spectra.
{\viibf We do not employ the Fast Fourier Transform (FFT) implementation
of the DFT. Hence, we use the light curves in their original form and do not
require rebinning to a regular grid, or special treatment of data gaps.}
The uncertainty on individual flux measurements in the light curves
can differ greatly due to the range in observation durations, and varied
location of sources in the BAT's field of view.
Because of this, the contribution of each
data point to the power spectrum is weighted by its uncertainty
using the ``semi-weighting'' technique \citep{Corbet2007a,Corbet2007b}, which takes into account
both the error bars on each data point and the excess variability of the
light curve. Weighting data points in a power spectrum
is analogous to combining individual data points using weighted means \citep{Scargle1989},
and semi-weighting is analogous to using semi-weighted means \citep{Cochran1937,Cochran1954}.
We calculated DFTs of the light curves for 
frequency ranges which correspond to
periods of between 0.07 days to the length of the light curves -
i.e. generally \sqig 4250 days.
We oversampled the DFTs by a factor of five compared to their nominal resolution,
{\viibf where
we take the nominal frequency resolution to be the inverse of the length of each light curve 
\citep[e.g.][and references therein]{VanderPlas2017}.}
{\viibf To estimate the significance of peaks in the power spectra we utilize the
commonly employed false alarm probability
\citep[FAP; ][]{Scargle1982}, while we note that possible statistical problems
with the FAP have been identified by several authors \citep[e.g.][]{Koen1990,Baluev2008,Suveges2014}.}
{\viibf We note that the calculation of the FAP can also be complicated if the underlying
continuum is not ``white''. Attempts to deal with this have employed e.g. the removal
of a smoothed continuum \citep{Israel1996} or polynomial fits to log(power) versus log(frequency) to
the entire power spectrum \citep[e.g.][]{Vaughan2005,Corbet2008}.}
{\viibf Here, to estimate FAP values of candidate periods in the presence of uncertain
noise properties of the overall variability, we use the local power level near
a peak in the power spectrum derived from a polynomial fit to a limited frequency range.}
{\viibf To calculate the FAP, the number of independent frequencies in the total frequency
range search should be known which depends on the frequency resolution. While this is not
precisely defined for unevenly sampled data \citep[e.g.][]{Koen1990}, we have previously
found that using the inverse of the light curve length provides a reasonable approximation
\citep{Corbet2008}.}
Uncertainties in periods are generally derived using the expression
of \citet{Horne1986}, {\viibf again using the local power level rather than the mean
of the total power spectrum.}
{\viibf
We do not apply a barycentric correction to the observation times 
for the analysis presented here. In all cases, barycentric corrections
would be much smaller than uncertainties on the periods considered.
}

A DFT based approach to period searching is most sensitive to sinusoidal
modulation. Modulation such as a repetitive flare that is limited to
a brief orbital phase will result in the modulation being spread over
many harmonics in the power spectrum. While this can be alleviated by
summing a certain number of harmonics \citep[e.g.][]{Buccheri1983}{\mybf ,} we do
not generally see {\mybf sharp} flaring in our sources and do not employ this here.
While period searches based on
folding may be more sensitive to brief flare-like behavior,
we note though that folding techniques can result in
spurious signals at ``sub-harmonics'' that do not occur
in Fourier-based analyses.

\subsection{Period Persistence Tests: Dynamic Power Spectra, Time-Slice Folds,
and Height vs. Time Plots}

To investigate the persistence of a possible signal, it can be illustrative
to investigate changes in the power spectrum of a light
curve as a function of time. In this case, power spectra are
calculated of temporal subsets of the light curve, potentially
with overlap between subsets. Images of the power spectrum
as a function of time can then be instructive about changes in
modulation, and appearance and disappearance of signals.
However, the periods of the sources discussed here are
relatively long compared to the total light curve length,
and may also require a light curve subset that is a significant
fraction of the total length in order to be detected.
For the dynamic power spectra, it is therefore often required
to use overlapping light curve subsets that are not statistically
independent.
The length of the temporal subsets, and the degree
of overlap, are not naturally defined and must be selected by a user.

We also employ simple folding of light curve subsets on periods obtained from
the entire light curve, but selecting temporal subsets of
the light curve which have no overlap. The length we choose for the subset
light curves varies from source to source, and depends on
factors including the strength of the modulation.

In addition, we can investigate the change in strength of
the modulation at a particular period as a function of time.
For a persistent coherent modulation with constant level of modulation,
an approximately linear change in peak power relative to
the average power with time is expected.
Changes in the gradient of the change in signal strength 
can thus reveal times when signal strength was relatively stronger or
weaker.

\newpage
\begin{figure}
\epsscale{0.8}
\includegraphics[width=7.25cm]{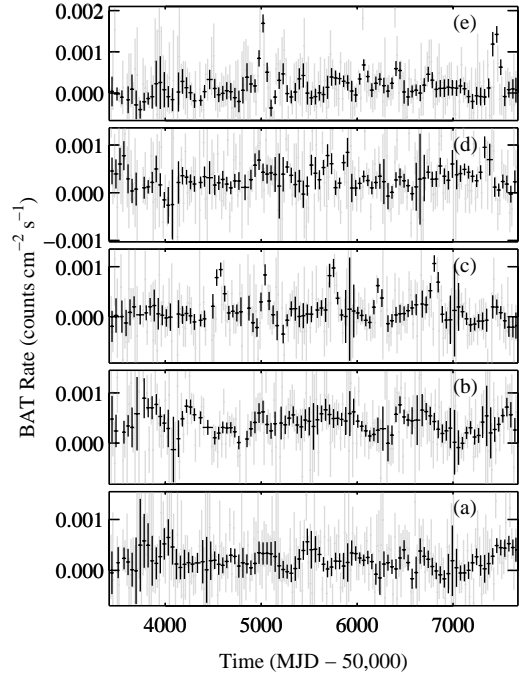}
\caption{{\newbf BAT light curves of: (a) \igr1448, (b) AX J1700.2-4220, 
(c) Swift J1816.7-1613, (d) AX J1820.5-1434, and (e) XTE J1906+090.
The gray points show light curves derived from the one day average light curves,
rebinned to a time resolution of 21 days. The black points 
are rebinned to a time resolution of 42 days, then smoothed by convolution with
a triangular response of full-width half maximum of 84 days.}
}
\label{fig:mega_lc}
\end{figure}
\newpage

\begin{figure}
\epsscale{0.8}
\includegraphics[width=13cm]{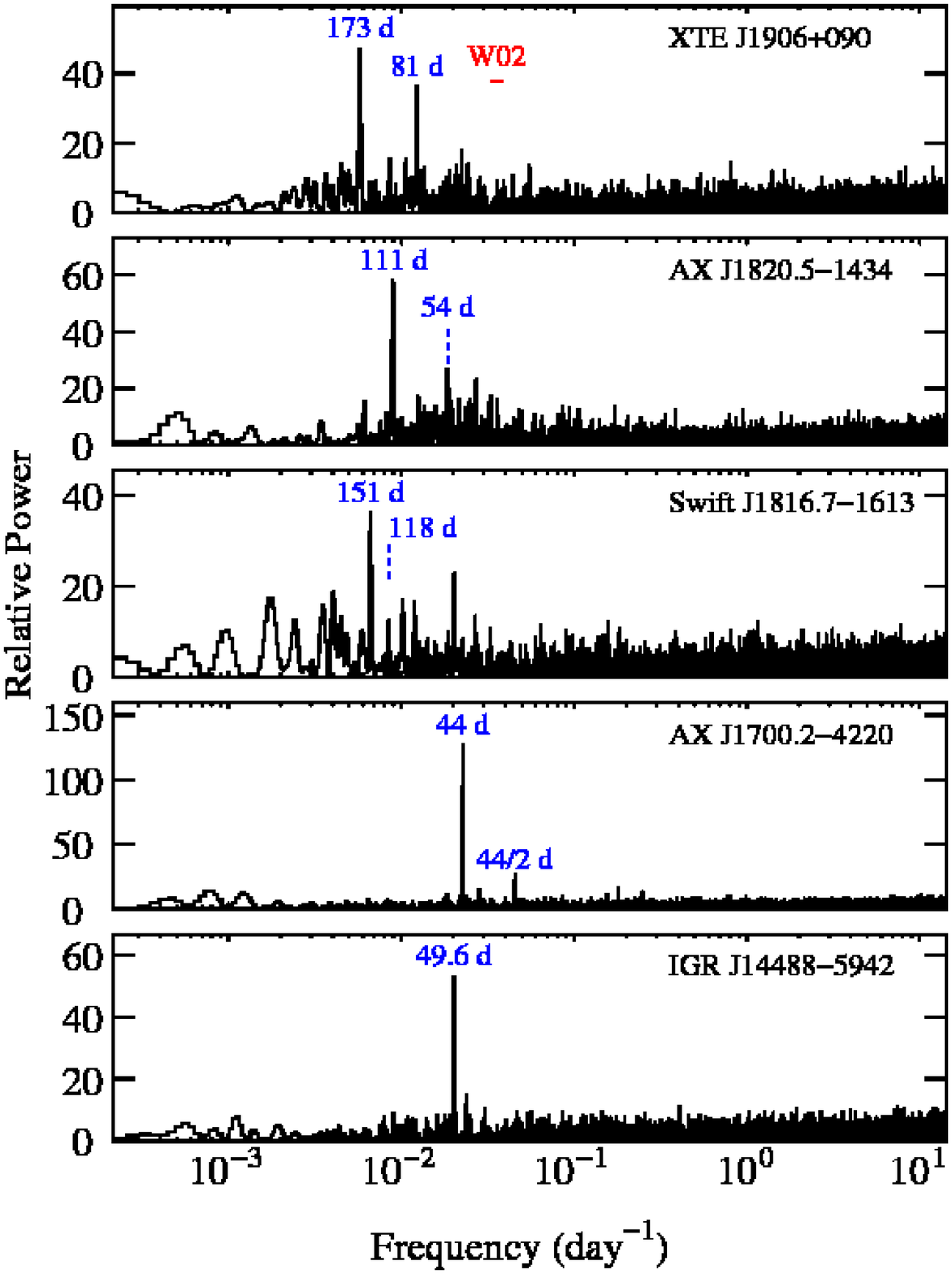}
\caption{{\mybf Power spectra  of the BAT light curves of 
IGR J14488-5942, AX J1700.2-4220, 
Swift J1816.7-1613, AX J1820.5-1434, and XTE J1906+090.
For each source candidate periods are marked. 
For Swift J1816.7-1613 the dashed lines mark the 118 and 151 day periods reported by
\citet{LaParola2014} and \citet{Corbet2014} respectively. 
For AX J1820.5-1434 the dashed line marks the 54 day period reported by \citet{Segreto2013}.
The highest peak in the power spectrum is at a period of \sqig111 days. For
XTE J1906+090 the two highest peaks at \sqig173 and 81 days
are marked. The short horizontal line marked ``W02'' is the 26--30
day period range suggested by \citet{Wilson2002} from pulse timing.
See text for additional details.}
}
\label{fig:bat_power}
\end{figure}

\clearpage

\section{{\viibf Sources and Analysis Results}}

\subsection{\igr1448}

\igr1448\ was first listed in the 4th INTEGRAL/IBIS Survey Catalogue 
\citep{Bird2010} as a variable source. Although it was not present in the BAT 22 month survey 
\citep{Tueller2010}, it is present in the 70 month catalog \citep{Baumgartner2013}
as Swift J1448.4-5945.
\citet{Landi2009} performed \Swift\ XRT observations covering 
the location of \igrj1448\ and found two sources: ``N1" and ``N2". They proposed that the 
brighter and spectrally harder source, N2 (Swift J144843.3-594216), was the counterpart of \igrj1448. 
\citet{Rodriguez2010} also analyzed the \Swift\ XRT observations and {\newbf also} slightly preferred 
the brighter source as
the counterpart, in which case it could be a highly-absorbed X-ray binary.
From optical spectroscopy, \citet{Coleiro2013} suggested that the system
is more likely to contain a Be star {\newbf primary} than a supergiant.

\subsubsection{BAT Observations of \igr1448}

\begin{figure}
\epsscale{0.8}
\includegraphics[width=7.25cm,angle=270]{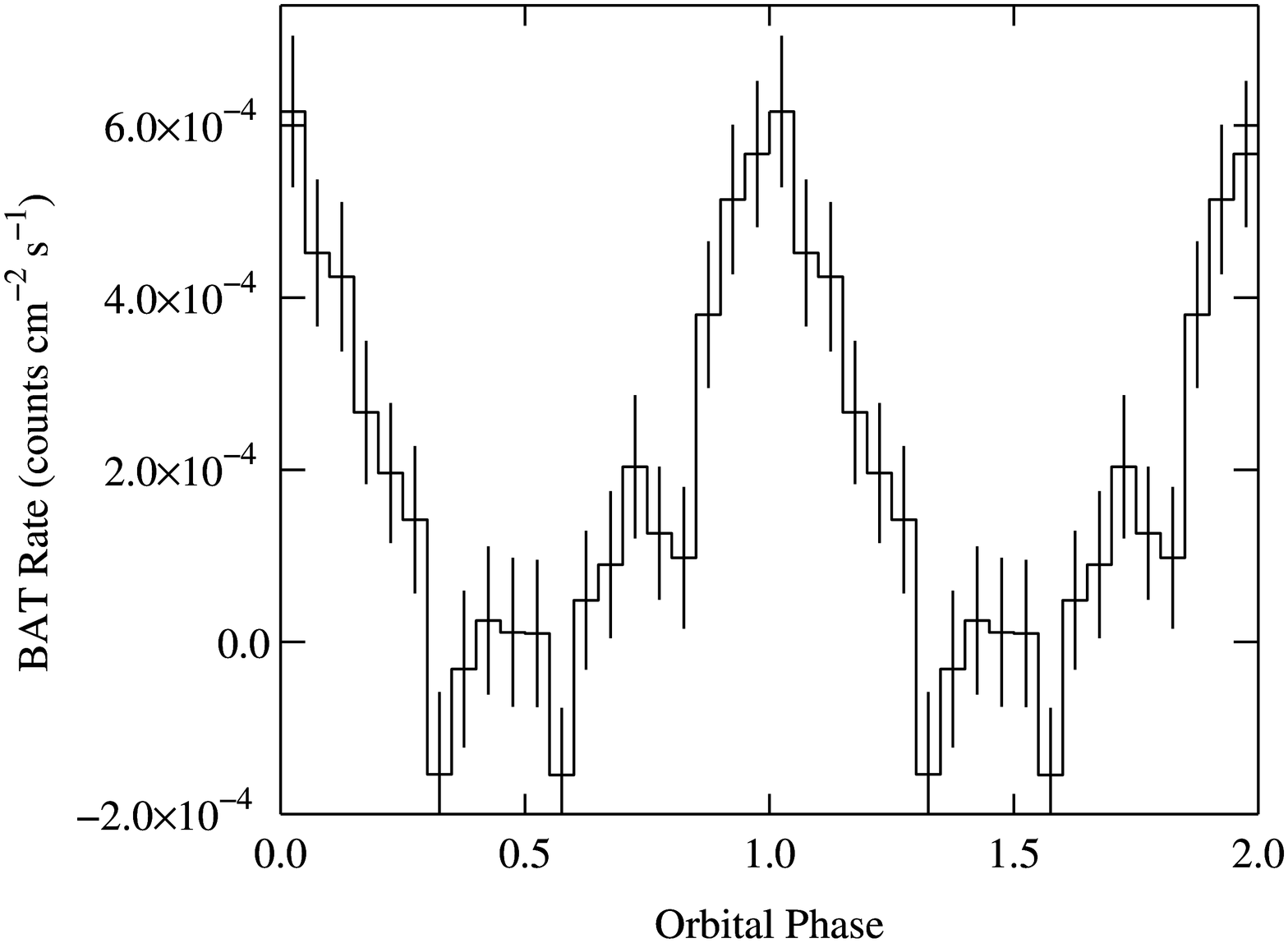}
\caption{BAT light curve of \igr1448
folded on the proposed {\newbf 49.63} day orbital period.
{\mybf Phase 0 corresponds to MJD 56,054.5.}
}
\label{fig:1448_fold}
\end{figure}

\begin{figure}
\epsscale{0.8}
\includegraphics[width=7.25cm,angle=270]{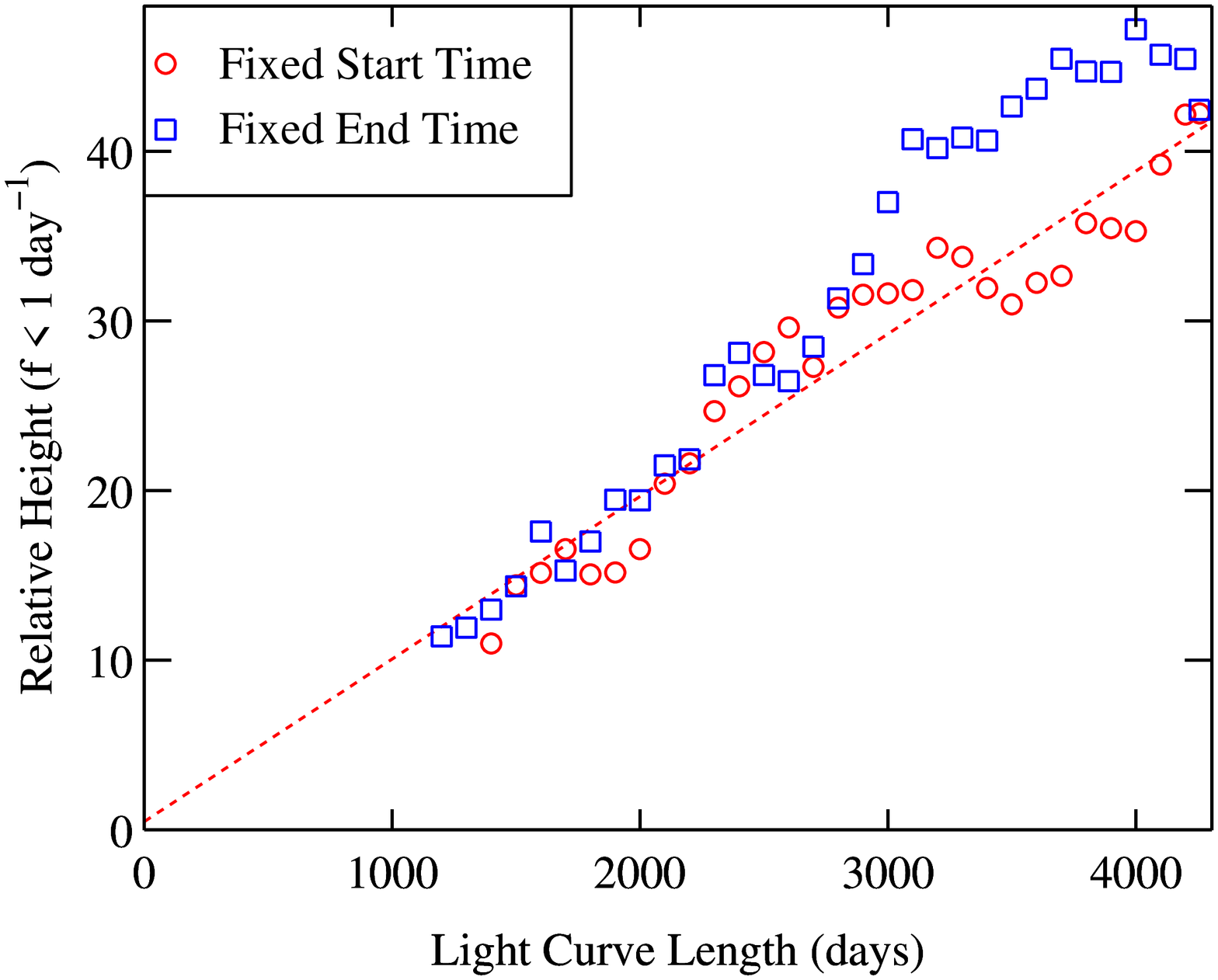}
\caption{Relative height of the peak near 49.6 days in the
power spectrum of the BAT light curve of \igr1448\ as a function of light
curve length. Red circles indicate light curves which all
have the same start time {\mybf(MJD 53,416)}, and blue squares are light curves
with the same end time {\mybf(MJD 57,673)}. 
}
\label{fig:1448_rvt}
\end{figure}

The BAT light curve of \igr1448\ is shown in Figure \ref{fig:mega_lc}\,(a).
{\viibf The mean count rate is (1.9 $\pm$ 0.2) $\times$10$^{-4}$\,\ctscm2s (\sqig0.9 mCrab).}
The power spectrum of the light curve for periods longer than 0.07 days
is shown in Figure \ref{fig:bat_power}.
A single highly significant peak is seen near 49.6 days.
The false alarm probability is $<$ 10$^{-6}$.
The period is 49.63 $\pm$ 0.05 days.
This is consistent with, but more precise than{\mybf ,} the value of 49.51 $\pm$ 0.12 d given
in \citet{Corbet2010b}.
The folded light curve is shown in Figure \ref{fig:1448_fold},
and it shows an approximately sinusoidal modulation.
From a sine wave fit to the light curve we derive an
epoch of maximum flux of MJD {\mybf56,054.5} $\pm$ 0.7.
The change of the peak height at 49.63 days in the power spectrum shows a fairly
constant increase with time (Figure \ref{fig:1448_rvt})
suggesting that modulation at this period is a persistent
property of the light curve.

\subsubsection{\Swift\ XRT Observations of \igrj1448}

\begin{figure}
\epsscale{0.8}
\includegraphics[width=7.25cm,angle=0]{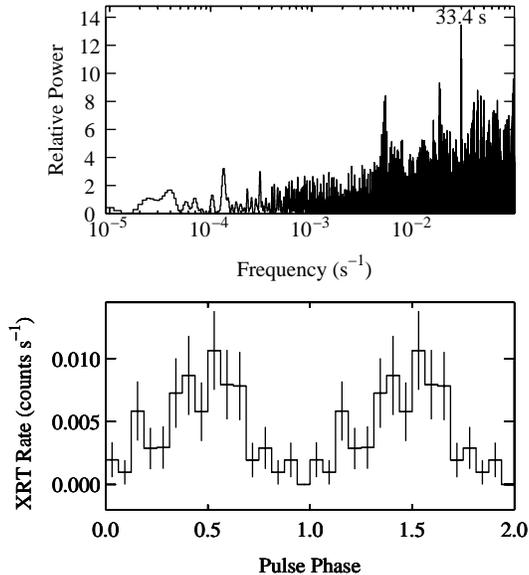}
\caption{
{\newbf
Top: Power spectrum  of the Swift XRT light curve of \igr1448\ {\newbf possible}
counterpart {\newbf Swift J144900.5-594503}.
Bottom: XRT light curve of {\newbf Swift J144900.5-594503}
folded on the possible 33.4 s pulse period.
}
}
\label{fig:xrt_n2_power_fold}
\end{figure}

{\newbf
The \Swift\ XRT observations of \igrj1448 described by
\citet{Landi2009} and \citet{Rodriguez2010} were obtained on 2009-09-25 (MJD 55,099)
with an exposure time of \sqig16.4 ks. Two sources were detected in the region,
the brighter source is Swift J144843.3-594216 (``Src \#1'' in \citealt{Rodriguez2010},
``N2'' in \citealt{Landi2009}) and the fainter source is Swift J144900.5-594503
(``Src \#2'' in \citealt{Rodriguez2010}, ``N1'' in \citealt{Landi2009}).
The time of the XRT observations corresponds to a phase of 0.93 when \igrj1448\
would be expected to be bright. 
The XRT data were obtained in photon-counting (PC) mode which has a time resolution of 2.5 s.
We extracted XRT light curves using aperture photometry, with aperture radii of 1.25\arcmin, 
{\mybf following the procedures described
in the \Swift\ XRT Data Reduction Guide \citep{Capalbi2005}.
For the brighter source, Swift J144843.3-594216, the non-background subtracted count rate in the aperture was 0.25 \cts\
and for the fainter source, Swift J144900.5-594503, it was 0.004 \cts. 
We searched for pulsations for periods longer than 10 s in both sources. 
No pulsations were seen in Swift J144843.3-594216. 
For the fainter source, Swift J144900.5-594503, a marginal signal was seen at a period of 33.419 $\pm$ 0.001 seconds
with an FAP of \sqig1.4\% (Figure \ref{fig:xrt_n2_power_fold}). The XRT light curve folded on this
period is also shown in Figure \ref{fig:xrt_n2_power_fold}.
}}

\subsection{\axj1700}

AX J170017-4220 was included in a list of faint X-ray sources found
in a survey of the Galactic plane with ASCA \citep{Sugizaki2001}.
This source is then listed, as \ax1700, as a counterpart to 
a source in the first INTEGRAL catalog \citep{Bird2004}.
From CTIO 1.5 m telescope optical spectroscopy,
\citet{Masetti2006} tentatively identified the Be star HD 153295, which
lies within the INTEGRAL error region, but outside the smaller ASCA error region, as the counterpart.
\citet{Negueruela2007} presented intermediate-resolution optical spectroscopy{\mybf ,} obtained
with the 1.9m telescope at the South African Astronomical Observatory{\mybf ,} and derived
a spectral type of B0.5 IVe.
\Chandra\ observations by \citet{Anderson2014} confirmed the identification of \ax1700,
via the detection of a source which they termed ChI J170017–4220\_1, with HD 153295.
\citet{Anderson2014} also noted that HD 153295 had unusual near-infrared colors.

\subsubsection{BAT Observations of \axj1700}

\begin{figure}
\epsscale{0.8}
\includegraphics[width=7.25cm,angle=270]{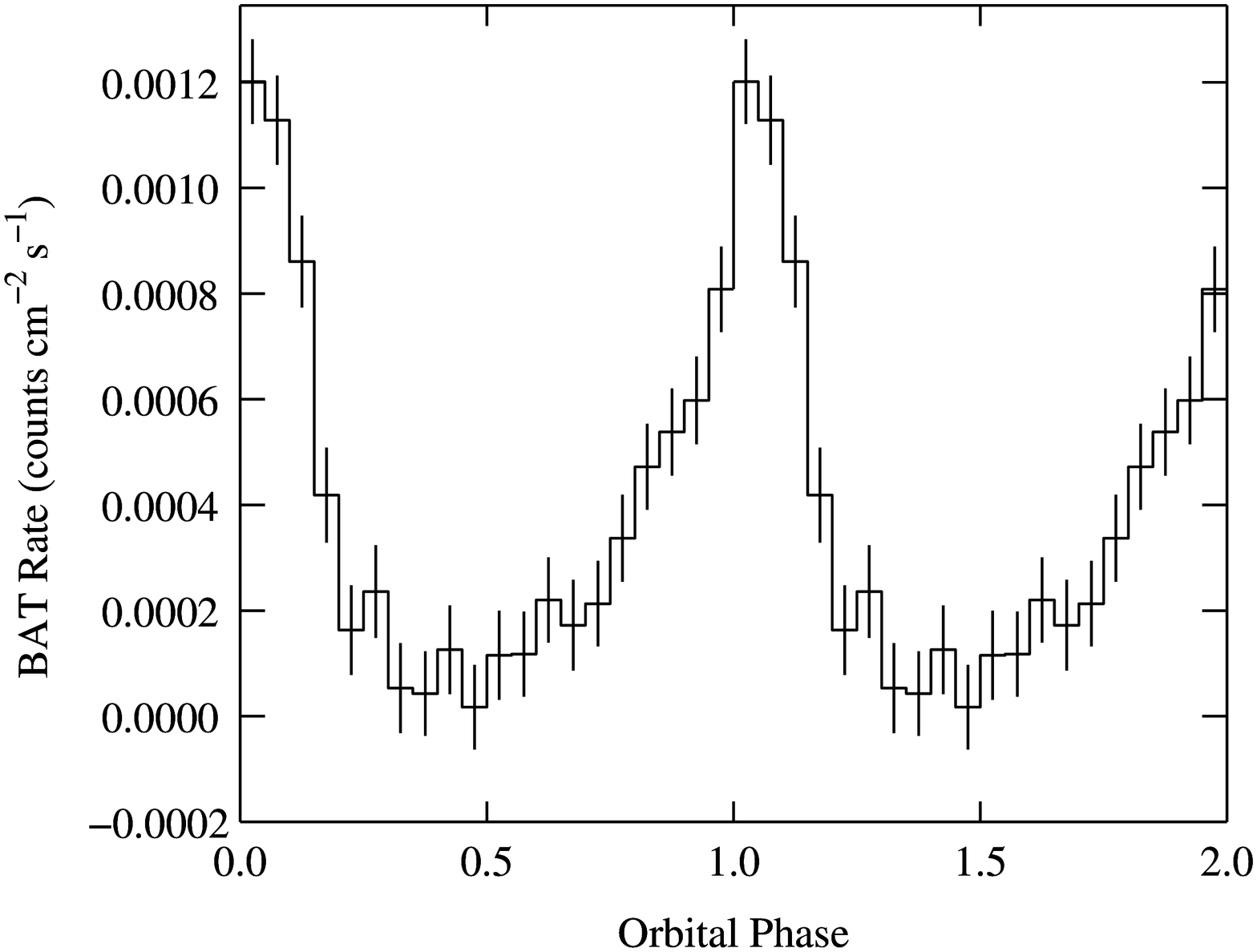}
\caption{BAT light curve of \axj1700\
folded on the proposed {\newbf 44.03} day orbital period.
Phase 0 corresponds to MJD 56,059.2.
}
\label{fig:1700_fold}
\end{figure}

The BAT light curve of \axj1700\ is shown in {\newbf Figure \ref{fig:mega_lc}\,(b)}.
{\viibf The mean count rate is (3.9 $\pm$ 0.2) $\times$10$^{-4}$\,\ctscm2s (\sqig1.8 mCrab).}
The power spectrum of the light curve is shown in Figure \ref{fig:bat_power}.
A highly significant peak is seen near 44 days, together with a significant
peak at the second harmonic of this.
The false alarm probability is $<$ 10$^{-6}$ {\mybf and}
the period is 44.03 $\pm$ 0.03 days. {\newbf This period is
consistent at the \sqig2 $\sigma$ level with, but more precise than,} the value of 44.12 $\pm$ 0.04 days given
in \citet{Corbet2010a}.
The folded light curve is shown in Figure \ref{fig:1700_fold}, and
the folded profile is roughly sinusoidal, although perhaps
with a somewhat slower rise to maximum than subsequent decline.
From a sine wave fit to the light curve, we derive an
epoch of maximum flux of MJD 56,059.2 $\pm$ 0.4.

\begin{figure}
\epsscale{0.8}
\includegraphics[width=7.25cm,angle=270]{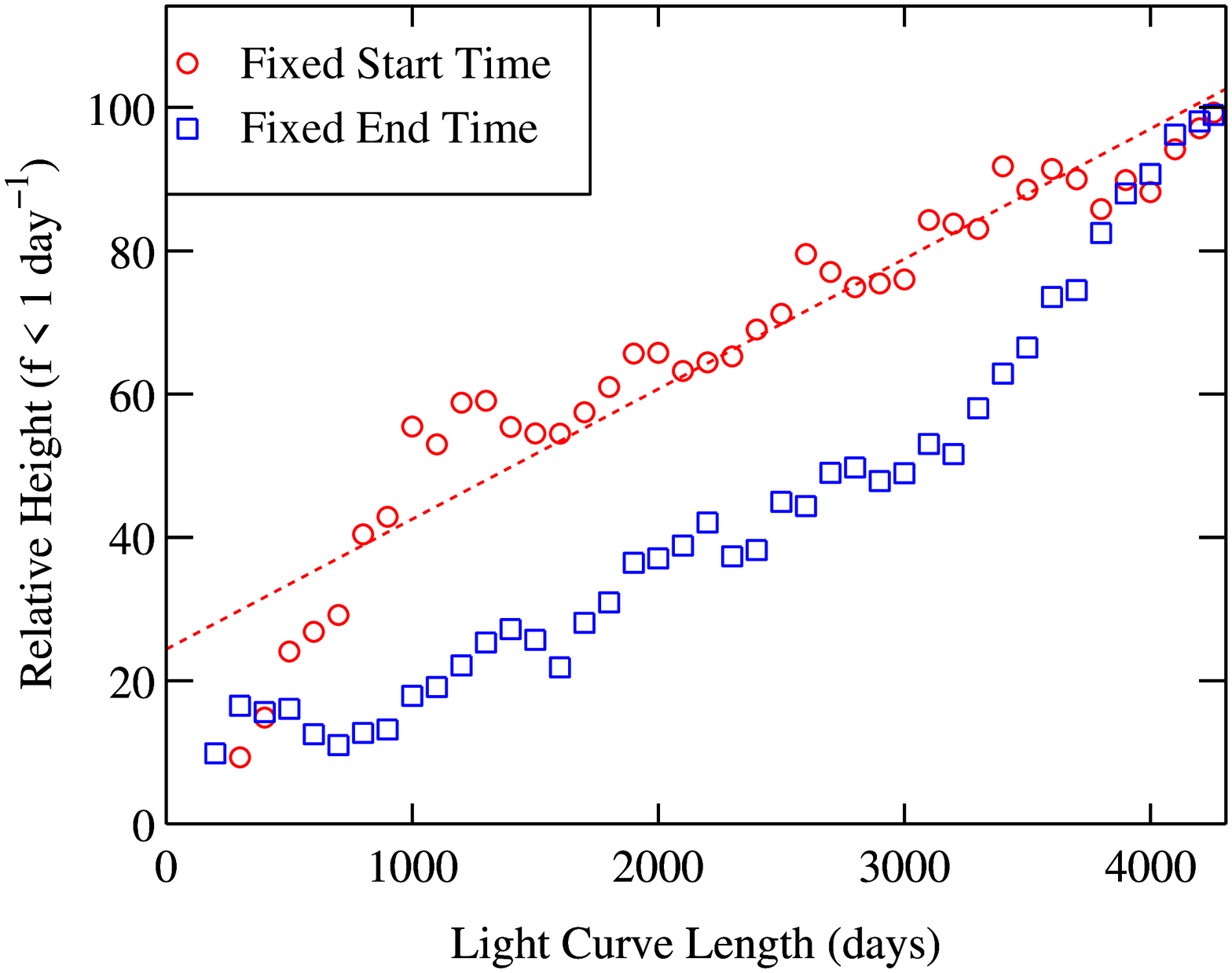}
\caption{Relative height of the peak near 44 days in the
power spectrum of the BAT light curve of \axj1700\ as a function of light
curve length. Red circles indicate light curves which all
have the same start time \mybf{(MJD 53,416)}, and blue squares are light curves
with the same end time \mybf{(MJD 57,673)}. 
}
\label{fig:1700_rvt}
\end{figure}

\begin{figure}
\epsscale{0.8}
\includegraphics[width=7.25cm,angle=0]{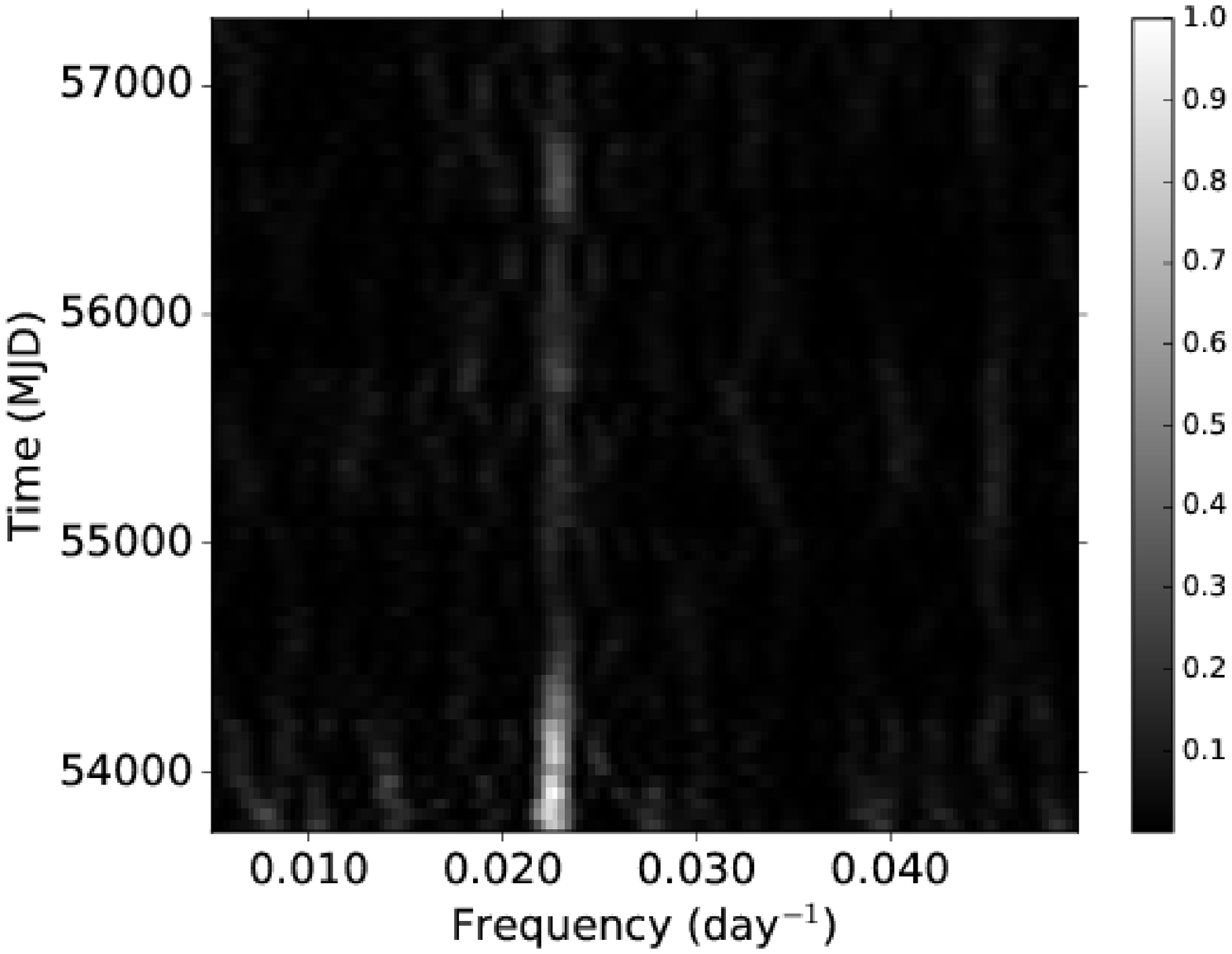}
\caption{Power spectrum of the BAT light curve of \axj1700\ as a function of time.
The low and high frequencies correspond to periods of 200 and 20 days, respectively.
The power spectra were calculated for light curve segments of length 650 days, with
increments in start and end times of the segments of 50 days.
{\viibf
The power spectra are normalized to the maximum power at any time.
}
}
\label{fig:1700_2dps}
\end{figure}

\begin{figure}
\epsscale{0.8}
\includegraphics[width=7.25cm,angle=0]{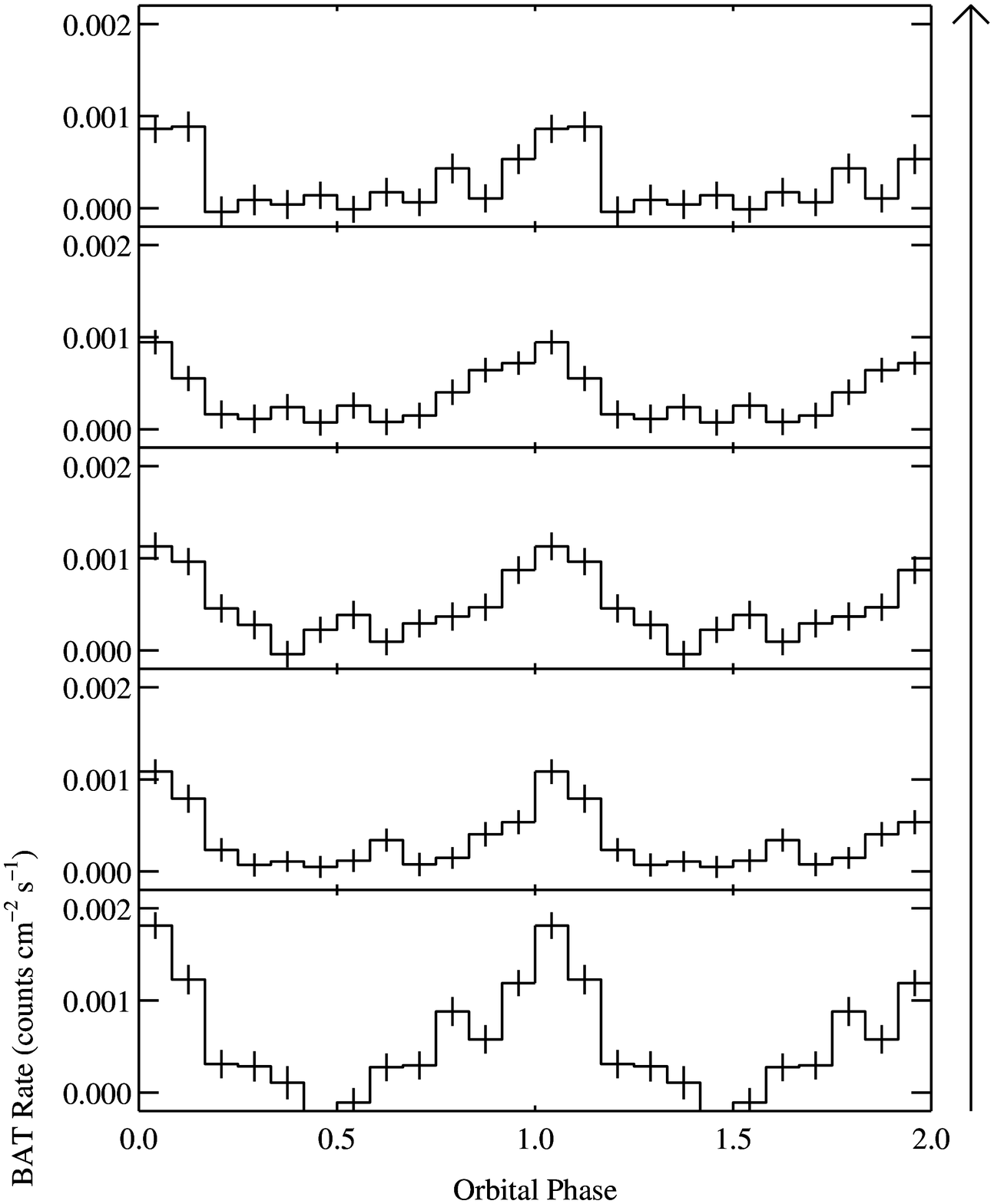}
\caption{Light curve of BAT light curve of \axj1700\ divided into
equal length {\mybf(851 day)} sections, and each folded on the {\newbf 44.03} day period. The arrow
indicates increasing time.
{\mybf Phase 0 corresponds to MJD 56,059.2.}
}
\label{fig:1700_slice_fold}
\end{figure}

The change of the peak height near 44 days in the power spectrum (Figure \ref{fig:1700_rvt}){\mybf ,}
shows an initially faster increase with increasing light curve
length than for later in the light curve. This suggests that, although
modulation near 44 days is a persistent property of the light curve,
it was more prominent for the first \sqig 1000 days (MJD 53,416 to \sqig 54,400).
The dynamic power spectrum of the BAT light curve (Figure \ref{fig:1700_2dps}) is also consistent with
the modulation being stronger for the earlier portion of the light curve.
We divided the entire light curve into five equal length sections and folded each
on the 44 day period, and this is shown in Figure \ref{fig:1700_slice_fold}.
This shows that the modulation persists throughout the light curve, although the
source is brightest and shows the strongest modulation at earliest times.

\subsubsection{\RXTE\ Observations of \axj1700}
\begin{figure}
\epsscale{0.8}
\includegraphics[width=7.25cm,angle=0]{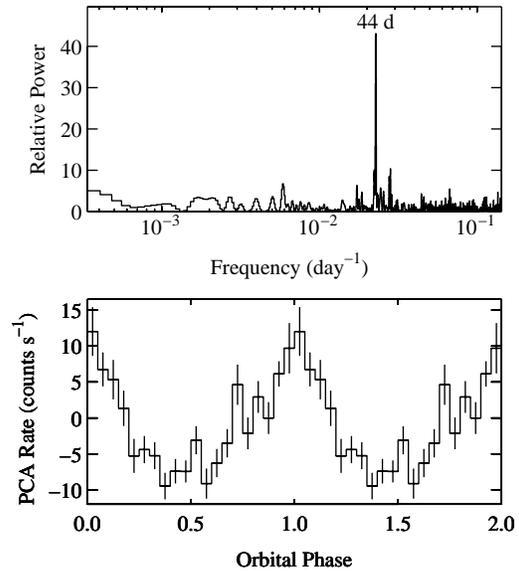}
\caption{{\newbf Top panel: Power spectrum  of the PCA light curve of \ax1700.
Bottom panel: PCA light curve of \ax1700
folded on the proposed {\newbf 44.03} day orbital period.
Phase 0 corresponds to MJD 56,059.2.}
}
\label{fig:1700_pca_power_fold}
\end{figure}

Long-term observations of sources near the Galactic center are available
from scans across this region made with the \RXTE\ PCA \citep{Swank2001}{\mybf ,}
and cover an energy range of 2 - 10 keV.
Detection of the orbital period of \axj1700\ from \RXTE\ PCA Galactic plane
observations was reported by \citet{Markwardt2010}{\mybf ,} who obtained a period
of 44.03 $\pm$ 0.14 days.
We investigate here the full PCA light curve that covers
the time range 2004-06-10 (MJD 53,166) to 2011-10-29 (MJD 55,863){\mybf ,} which is \sqig 25\% longer than
that available to \citet{Markwardt2010}.
The power spectrum of the PCA light curve is shown in Figure \ref{fig:1700_pca_power_fold}.
The orbital period derived from the PCA scan light curve is
43.98 $\pm$ 0.07 days, which is consistent with the period derived with the BAT.
The {\newbf PCA} light curve folded on the presumed orbital period is also shown in
Figure \ref{fig:1700_pca_power_fold}. This appears similar to the folded
BAT light curve. The lowest flux counts have negative values, which are
likely due to the flux bias noted by \citet{Markwardt2010}.

We also examined the 1.5 - 12 keV \RXTE\ All-sky Monitor \citep{Levine1996} light curve of \axj1700
which covers the time range MJD 50,088 to 55,813 (1996-01-06 to 2011-09-09).
The power spectrum of this shows no evidence for modulation
on the 44 day period.

\subsection{BAT Observations of Swift J1816.7-1613}

The transient X-ray source Swift J1816.7-1613 was discovered with the 
\Swift\ BAT by \citet{Krimm2008}. 
A pulse period of 143 s was found by \citet{Halpern2008} from 
Chandra observations, and this was confirmed by \citet{Krimm2013} from \RXTE\
PCA data. 
Analysis of archival BeppoSAX data also indicated an earlier detection of
the source \citep{Orlandini2008}.
\citet{Corbet2014} proposed a possible orbital period of 151.4 $\pm$ 1 days using
\Swift\ BAT data.
However, \citet{LaParola2014} reported the detection of a 118.5 $\pm$ 0.8 day period, also from BAT
data, but using only data obtained around the time of apparent flares in the light curve.
No optical counterpart has so far been identified. However, \citet{Corbet2014}
and \citet{LaParola2014} both suggested a Be star classification based
on the location of the source in the spin/orbital period diagram \citep{Corbet1986} for their proposed
orbital periods.

\begin{figure}
\epsscale{0.8}
\includegraphics[width=7.25cm,angle=270]{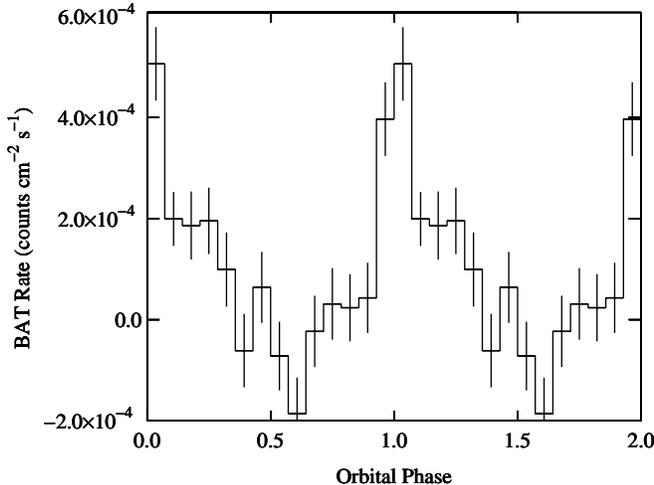}
\caption{BAT light curve of Swift J1816.7-1613
folded on a period of 151.1 days.
{\mybf Phase 0 corresponds to MJD 56,967.}
}
\label{fig:1816_fold}
\end{figure}

\begin{figure}
\epsscale{0.8}
\includegraphics[width=7.25cm,angle=270]{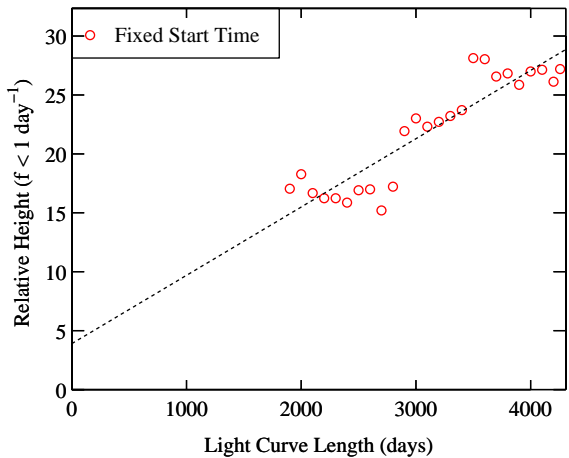}
\caption{Relative height of the peak near 151 days in the
power spectrum of the BAT light curve of Swift J1816.7-1613 as a function of light
curve length {\mybf with all light curves having a start time of MJD 53,416}. 
}
\label{fig:1816_rvt}
\end{figure}

\begin{figure}
\epsscale{0.8}
\includegraphics[width=7.25cm,angle=0]{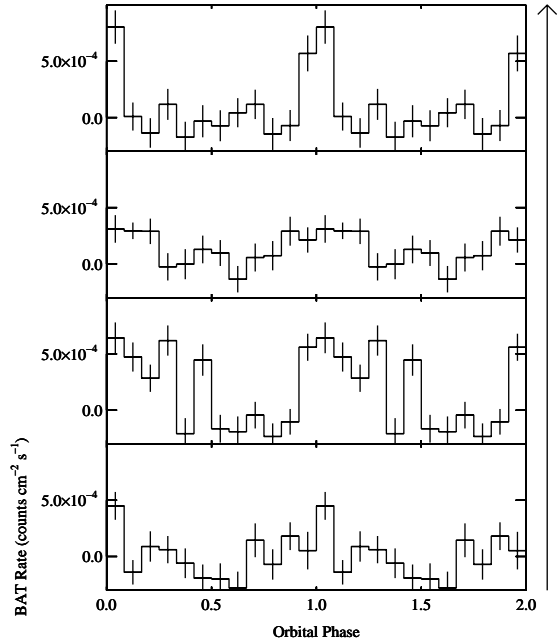}
\caption{Light curve of BAT light curve of Swift J1816.7-1613 divided into
equal length {\mybf(1064 day)} sections, and each folded on the possible {\newbf 151.1} day period. The arrow
indicates increasing time.
{\mybf Phase 0 corresponds to MJD 56,967.}
}
\label{fig:1816_slice_fold}
\end{figure}

The BAT light curve of \swiftj1816\ is shown in Figure \ref{fig:mega_lc}\,(c),
the power spectrum of this is shown in Figure \ref{fig:bat_power}.
The mean count rate is (1.1 $\pm$ 0.2) $\times\,$10$^{-4}$\,\ctscm2s (\sqig0.5 mCrab).
The strongest peak is at a period of 151.1 $\pm$ 0.5 days{\mybf ,} which is consistent with the
151.4 $\pm$ 1 day period reported in \citet{Corbet2014}.
Although the peak height is \sqig36 times larger than the mean power level, the
noise level near this peak appears higher. Using a {\newbf 2nd order} polynomial fit to obtain the local
power, the relative height is \sqig8 times the local power level, and so is not
statistically significant. 
The 118 day period reported by \citet{LaParola2014}, based on selected subsets
of the light curve,
is not detected in the power spectrum of the entire light curve.
In Figure \ref{fig:1816_fold} we show the BAT light curve folded
on the 151.1 period, with an assumed epoch of maximum flux of
MJD 56,967. This shows an apparently fairly sharply peaked profile.
A plot of height of the peak in the power spectrum at this
period with time (Figure \ref{fig:1816_rvt}) shows that while there is a general trend to increase with time,
the height does not increase consistently.
Folding sections of the BAT light curve on the 151.1 day period
(Figure \ref{fig:1816_slice_fold}),
shows that
the modulation is most clearly seen in the most recent quarter (1064 days) 
of the light curve.

\begin{figure}
\epsscale{0.8}
\includegraphics[width=7.25cm,angle=270]{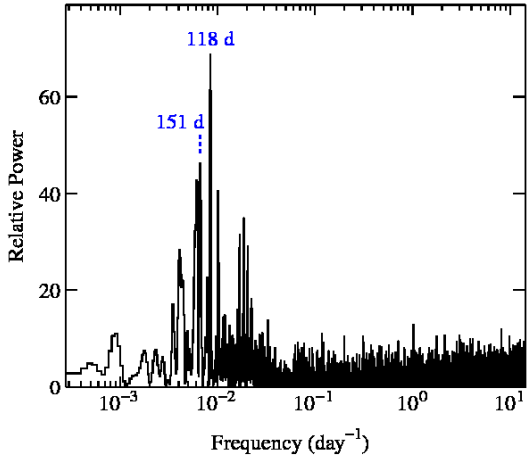}
\caption{Power spectrum of the BAT light curve of Swift J1816.7-1613 using only
the time ranges adopted by \citet{LaParola2014} around candidate flares.
}
\label{fig:laparola_power}
\end{figure}

\begin{figure}
\epsscale{0.8}
\includegraphics[width=7.25cm,angle=0]{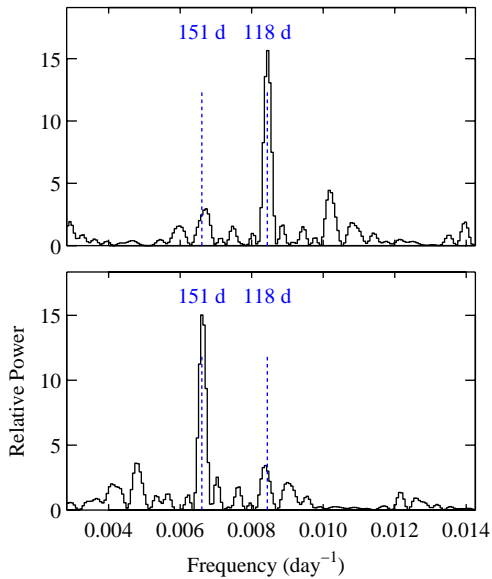}
\caption{{\viibf Power spectra of light curves with sine waves with periods
of 151.1 days (bottom) and 118.5 days (top) using data times from the BAT
light curve of Swift J1816.7-1613 using only
the time ranges adopted by \citet{LaParola2014} around candidate flares.
}}
\label{fig:1816_alias}
\end{figure}

We also calculated the power spectrum of \swiftj1816\ using only the restricted
time ranges adopted by \citet{LaParola2014} and this is shown in Figure \ref{fig:laparola_power}.
The maximum in the power spectrum is at a period of 118.9 $\pm$ 0.6 days which is consistent
with the period of 118.5 $\pm$ 0.8 days found by \citet{LaParola2014}.
We note that by selecting only time ranges around a modest number of
apparent flares, six gaps of length \sqig140 to 500 days are created in the light curve.
These gaps have the potential to causes ``aliases'' in power spectra and other
period search techniques. To investigate this, we created light curves
based on the BAT light curve times, but with the data values replaced
with sine wave modulations at 151.1 and 118.5 days. In both cases
aliasing created small peaks at the other period (Figure \ref{fig:1816_alias}).

\subsection{BAT Observations of AX J1820.5-1434}

AX J1820.5-1434 was discovered by \citet{Kinugasa1998} during a survey of the
Galactic plane using ASCA, and a pulse period of 152.26 $\pm$ 0.04 s was measured.
Searches for an optical counterpart have not yet resulted in a definite identification
\citep{Kaur2010, Negueruela2007}.
The source has been detected with \INTEGRAL\ \citep{Lutovinov2003,Lutovinov2005,Bird2010}
and, from flaring activity seen with this satellite, \citet{Walter2007} tentatively suggested
that the source was a Supergiant Fast X-ray Transient (SFXT).
\citet{Segreto2013} reported the determination of a 54.0 $\pm$ 0.4 day orbital
period for AX J1820.5-1434 from an analysis of \Swift\ BAT observations
obtained up to March 2012 (\sqig MJD 56,000)
and suggested
that this may be a Be star system. \citet{Segreto2013} also considered whether changes
in source variability might indicate that the recurrent flaring on the 54 day period
might {\it not} be related to periastron passages.
{\viibf The identification of the 54 day period by \citet{Segreto2013}
used a folding technique, and so sub-harmonics of their period were also
present in their periodogram.}

\begin{figure}
\epsscale{0.8}
\includegraphics[width=7.25cm,angle=270]{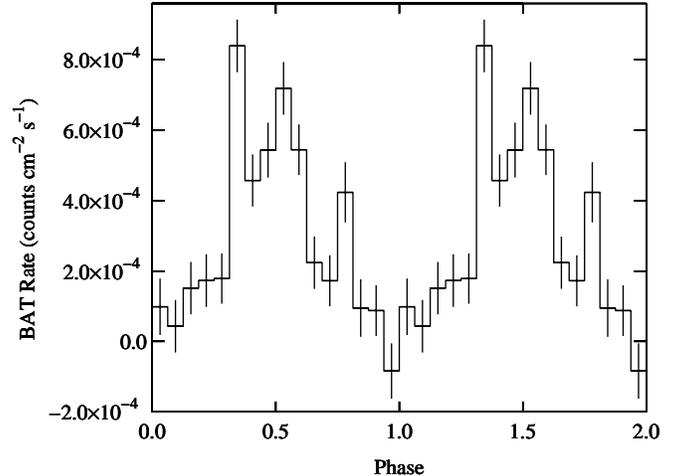}
\caption{BAT light curve of AX\,J1820.5-1434
folded on the possible {\newbf 111.8} day period.
Phase 0 corresponds to MJD 54,684.809 \citep{Segreto2013}.
}
\label{fig:1820_fold}
\end{figure}

The BAT light curve of AX J1820.5-1434 is shown in 
Figure \ref{fig:mega_lc}\,(d),
and the power spectrum of this
is plotted in
Figure \ref{fig:bat_power}.
The mean count rate is (2.9 $\pm$ 0.2) $\times$10$^{-4}$\,\ctscm2s (\sqig1.3 mCrab).
The strongest peak in the power spectrum is at a period of 111.8 $\pm$ 0.3 days.
While the peak is \sqig58 times the mean power level, it is only \sqig11
times the local power level.
The 111.8 day period is approximately, although formally inconsistent with,
twice the period of 54.0 $\pm$ 0.4 days reported by \citet{Segreto2013}.
The second strongest peak in the power spectrum has a much lower
amplitude and is at 54.1 $\pm$ 0.1 days. This small peak would be
consistent with the \citet{Segreto2013} period, but is not consistent
with the second harmonic of the 111.8 day period.
The BAT light curve folded on the 111.8 day period is shown in
Figure \ref{fig:1820_fold}.
We note that the profile, while having an overall smooth modulation,
is rather ``jagged'' with bin-to-bin variability.

\begin{figure}
\epsscale{0.8}
\includegraphics[width=7.25cm,angle=270]{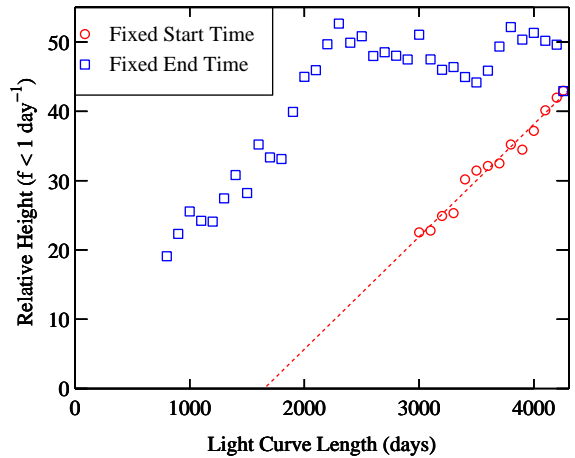}
\caption{Relative height of the peak near 111 days in the
power spectrum of the BAT light curve of AX\,J1820.5-1434 as a function of light
curve length. Red circles indicate light curves which all
have the same start time \mybf{(MJD 53,416)}, and blue squares are light curves
with the same end time \mybf{(MJD 57,673)}. 
}
\label{fig:1820_rvt}
\end{figure}

\begin{figure}
\epsscale{0.8}
\includegraphics[width=7.25cm,angle=0]{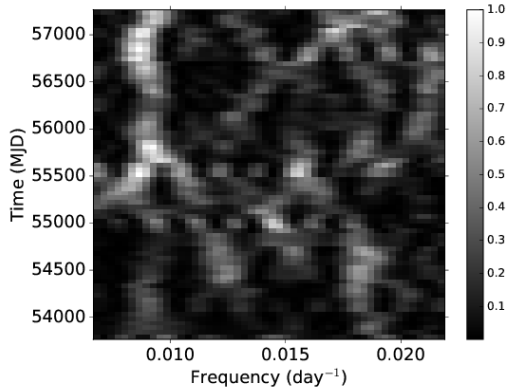}
\caption{Power spectrum of the BAT light curve of AX\,J1820.5-1434 as a function of time.
The low and high frequencies correspond to periods of 150 and 45 days, respectively.
The power spectra were calculated for light curve segments of length 700 days, with
increments in start and end times of the segments of 50 days.
The power spectra are normalized to the maximum power at any time.
}
\label{fig:1820_2dps}
\end{figure}

\begin{figure}
\epsscale{0.8}
\includegraphics[width=7.25cm,angle=0]{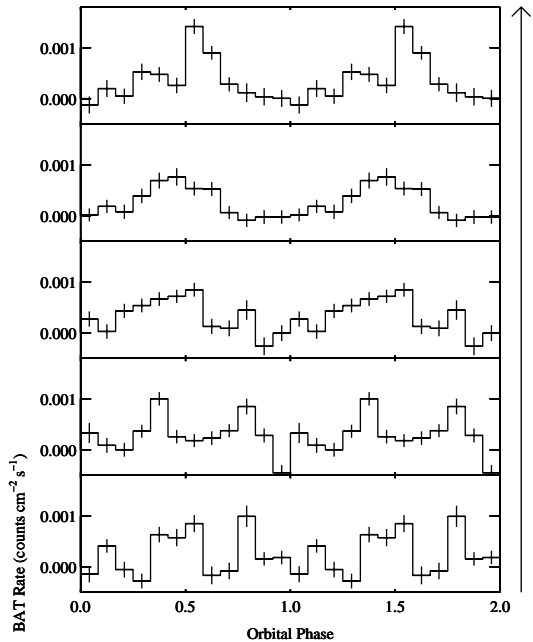}
\caption{Light curve of BAT light curve of AX\,J1820.5-1434   divided into
equal length {\mybf(851 day)} sections, and each folded on the {\newbf 111.8} day period. The arrow
indicates increasing time.
Phase 0 corresponds to MJD 54,684.809 \citep{Segreto2013}.
}
\label{fig:1820_slice_fold}
\end{figure}

From the plot of relative peak height near 111.8 days against time (Figure \ref{fig:1820_rvt}){\mybf ,} a
steady increase in peak height can be seen, but the
extrapolation back to a relative peak height of zero
starts at approximately 1650 days after the start of
the light curve (\sqig MJD 55,066). 
For a similar approach, but using increasing length light
curves that always {\it end} at the time of the last
observation (MJD 57,673) we find that the relative peak
height initially steadily increases as earlier start times
are used. However, the relative peak height reaches a maximum
then roughly plateaus for a light curve start time of \sqig2300
days before the end of the light curve. i.e. for times earlier
than \sqig MJD 55,370. This suggests that the periodic modulation
of the light curve became significantly stronger at some
time near MJD 55,200. 
The dynamic power spectrum of AX J1820.5-1434
(Figure \ref{fig:1820_2dps}) is also suggestive of modulation near
111.8 days (f \sqig0.009 day$^{-1}$) becoming apparent after \sqig MJD 55,200.
We show in Figure \ref{fig:1820_slice_fold} the light curve divided
into equal length sections folded on the 111.8 day period.
From this it can be seen that earlier time data shows a more strongly multi-peaked
profile than do later sections of the data.

\subsection{BAT Observations of XTE J1906+090}

{\newtext \citet{Marsden1998} serendipitously 
discovered the transient 89\,s pulsar XTE J1906+090 (= XTE J1906+09) 
during \RXTE\ PCA observations of the nearby source SGR 1900+14. 
From pulse timing{\mybf ,} \citet{Wilson2002} suggested an orbital period
in the range of 26--30 days. \citet{Wilson2002} noted that this combination of pulse and orbital
periods would make the system an outlier from the correlation between these periods found
for Be systems \citep{Corbet1984,Corbet1986}. From Chandra observations{\mybf ,} \citet{Gogus2005} identified
optical and infrared counterparts which they claimed, based on the optical and infrared
colors, added to the evidence that
XTE J1906+090 is a transient Be star X-ray binary.

The light curve of XTE J1906+090 is plotted in {\newbf Figure \ref{fig:mega_lc}\,(e),}
and the power spectrum of this
is shown in Figure \ref{fig:bat_power}. 
The mean count rate is (1.0 $\pm$ 0.2) $\times$10$^{-4}$\,\ctscm2s (\sqig0.5 mCrab).
No peak
is seen in the period range of 26 - 30 days suggested
by \citet{Wilson2002} from pulse timing.
Instead, the two strongest peaks are seen at periods of
173.1 $\pm$ 0.6 days and 81.4 $\pm$ 0.1 days.
We note that these periods are {\em not} consistent with being
harmonics of each other.
In Figures \ref{fig:1906_fold} and \ref{fig:1906_fold_2}
we show the BAT light curve folded on the 173.1 and 81.4 day
periods, respectively. The light curve folded on the 173.1 day
period shows sharply peaked modulation, while the data folded
on 81.4 days shows a smoother modulation.
In Figure \ref{fig:1906_81_173_rvt} we plot
the changes in relative peak heights as a function of increasing
light curve length. For both possible periods, although the tendency
is for peak height to increase with time, the increases are rather
erratic.

\begin{figure}
\epsscale{0.8}
\includegraphics[width=7.25cm,angle=270]{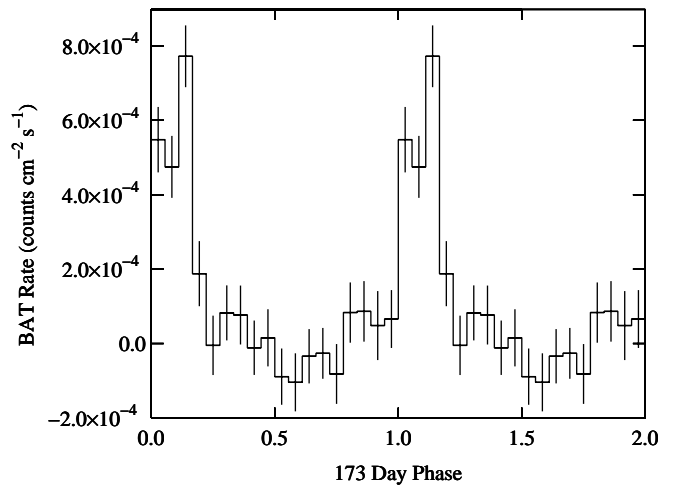}
\caption{BAT light curve of XTE J1906+090
folded on the {\newbf 173.1} day peak in the power spectrum.
Phase 0 corresponds to MJD 55,000.
}
\label{fig:1906_fold}
\end{figure}

\begin{figure}
\epsscale{0.8}
\includegraphics[width=7.25cm,angle=270]{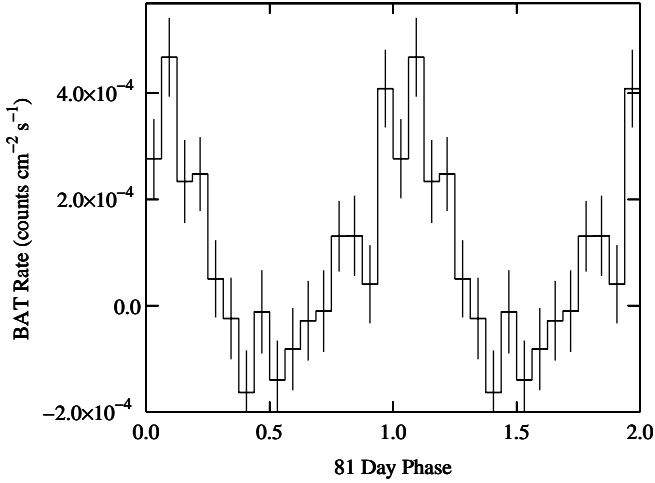}
\caption{BAT light curve of XTE J1906+090
folded on the {\newbf 81.4} day peak in the power spectrum.
Phase 0 corresponds to MJD 55,000.
}
\label{fig:1906_fold_2}
\end{figure}

In Figure \ref{fig:1906_phase_fold} we show the light curve
folded on the 81.4 day period with data selected on
two phase intervals of the 173 day period to correspond to
bright and faint phases of that possible modulation.
As would be expected, the data from the brighter phase
has a higher flux level. In addition, the modulation appears
stronger during the brighter phase{\mybf ,} with an apparent appearance
of a secondary peak offset by about 0.5 in phase from the primary
peak.

\begin{figure}
\epsscale{0.8}
\includegraphics[width=7.25cm]{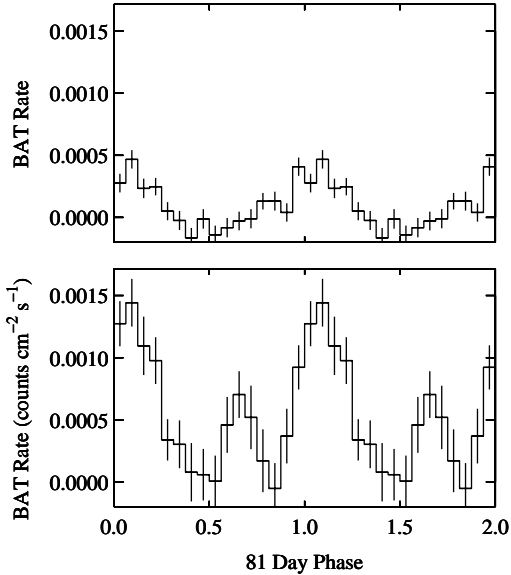}
\caption{BAT light curves of XTE J1906+090
folded on the 81.4 day peak in the power spectrum selected
by phase of the 173.1 day modulation. Top: phases 0.15 to 1.0.
Bottom: phases 0.0 to 0.15.
Phase 0 corresponds to MJD 55,000 for both the 81.4 and 173 day modulation.
}
\label{fig:1906_phase_fold}
\end{figure}

\begin{figure}
\epsscale{0.8}
\includegraphics[width=7.25cm,angle=0]{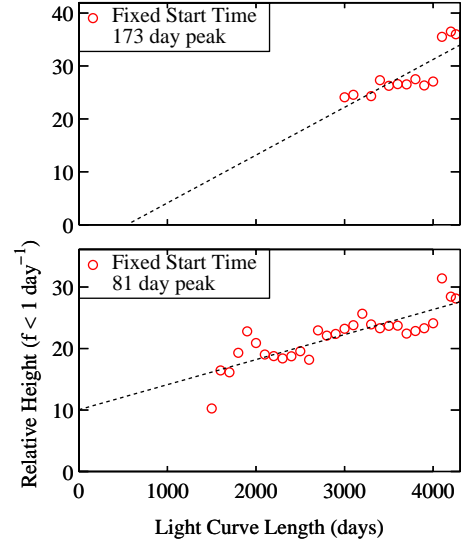}
\caption{{\newbf Bottom panel: Relative height of the peak near 81 days in the
power spectrum of the BAT light curve of XTE J1906+090 as a function of light
curve length, with all light curves having a start time of MJD 53,416.
Top panel: Relative height of the peak near 173 days in the
power spectrum of the BAT light curve of XTE J1906+090 as a function of light
curve length, with all light curves having a start time of MJD 53,416.}
}
\label{fig:1906_81_173_rvt}
\end{figure}

We note the presence in the long-term light curve {\newbf (Figure \ref{fig:mega_lc}\,(e))}
of two bright states/flares at approximately MJD 55,000 and {\mybf57,450}.
If data around these times are removed, then the power spectrum
still has a maximum at \sqig173 days, although at considerably
reduced power.

In Figure \ref{fig:1906_2dps} we show the dynamic power spectrum
of the BAT light curve for XTE J1906+090. The onset of modulation
at both \sqig 81 and 173 days appears to be associated with
the first flare in the light curve. 
{\newtext In Figures \ref{fig:1906_slice_81_fold} and \ref{fig:1906_slice_173_fold}
we show the light curve divided
into equal length sections folded on the 81.4 and 173.1 day periods, respectively.
These are also suggestive of increased modulation on these periods at later times.}

\begin{figure}
\epsscale{0.8}
\includegraphics[width=7.25cm,angle=0]{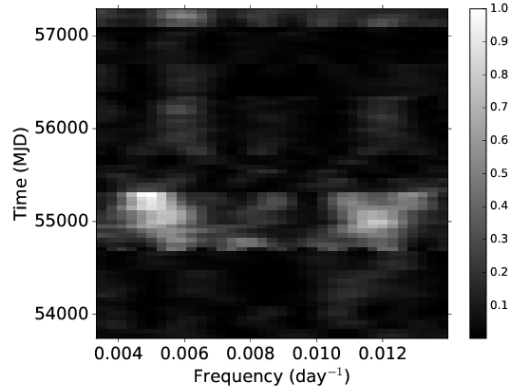}
\caption{Power spectrum of the BAT light curve of XTE J1906+090 as a function of time.
The low and high frequencies correspond to periods of 300 and 70 days, respectively.
The power spectra were calculated for light curve segments of length 650 days, with
increments in start and end times of the segments of 50 days.
{\viibf
The power spectra are normalized to the maximum power at any time.
}
}
\label{fig:1906_2dps}
\end{figure}

\begin{figure}
\epsscale{0.8}
\includegraphics[width=7.25cm,angle=0]{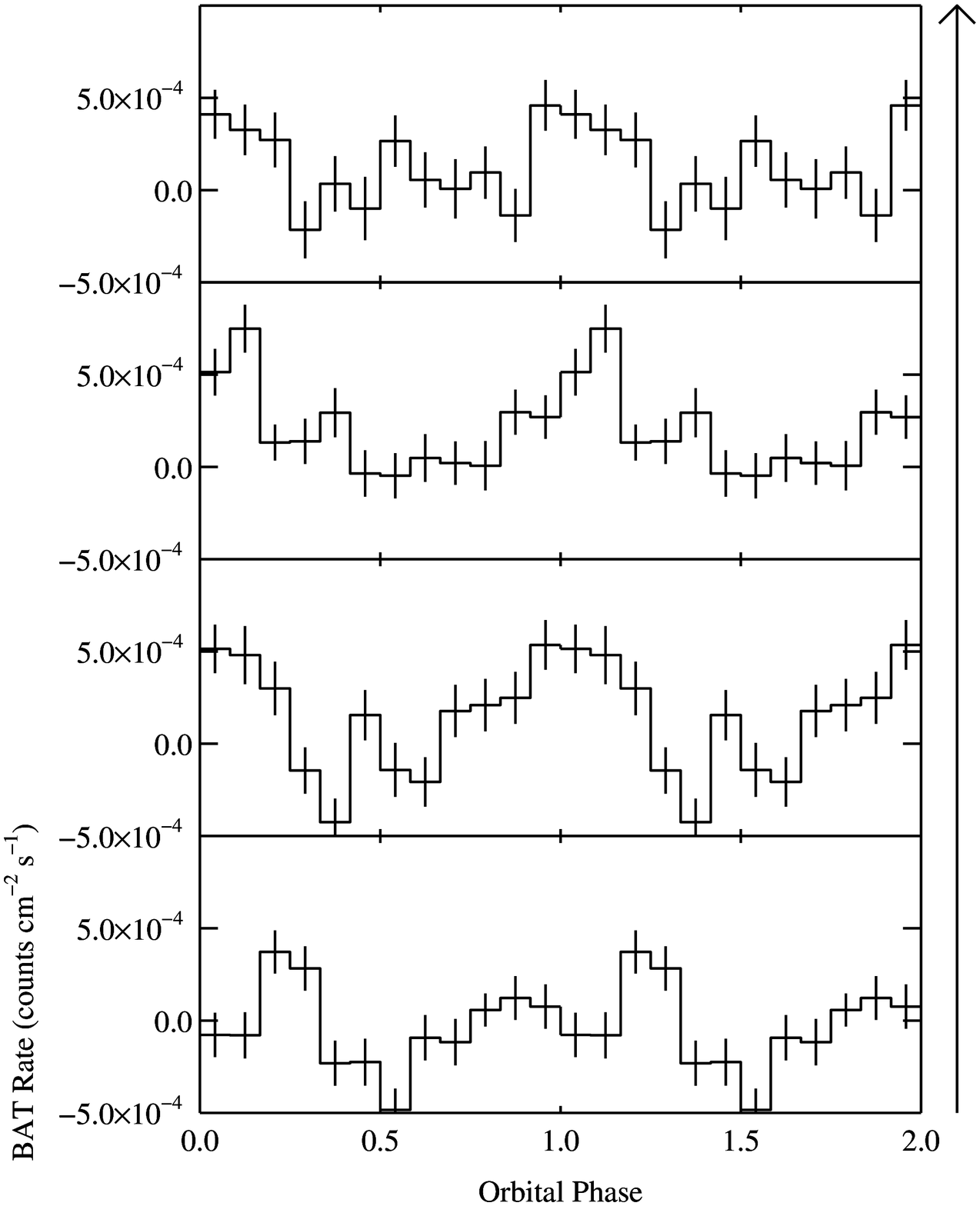}
\caption{Light curve of BAT light curve of XTE J1906+09 divided into
equal length (1064 day) sections, and each folded on the 81.4 day period. The arrow
indicates increasing time.
Phase 0 corresponds to MJD 55,000.
}
\label{fig:1906_slice_81_fold}
\end{figure}

\begin{figure}
\epsscale{0.8}
\includegraphics[width=7.25cm,angle=0]{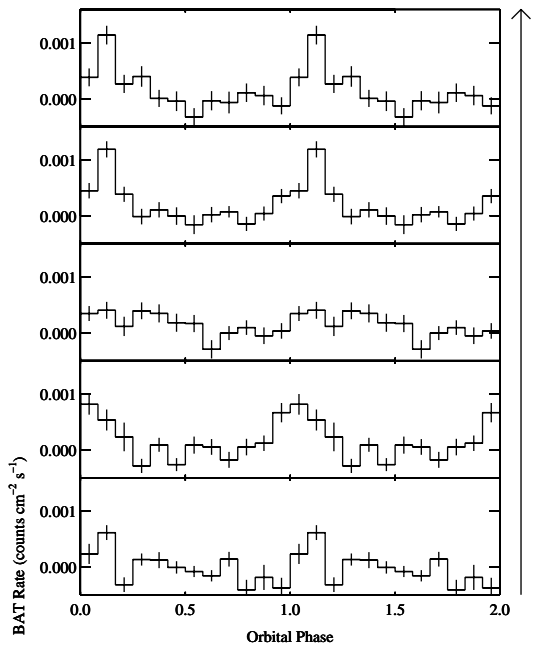}
\caption{Light curve of BAT light curve of XTE J1906+09 divided into
equal length {\mybf(851 day)} sections, and each folded on the {\newbf 173.1} day period. The arrow
indicates increasing time.
Phase 0 corresponds to MJD 55,000.
}
\label{fig:1906_slice_173_fold}
\end{figure}

\clearpage
\section{Discussion}

Even without an optical counterpart, the likely nature of the mass
donor in an HMXB can often be inferred from the length of the orbital period{\mybf ,} and
the nature of the orbital modulation. The inferences can generally
be stronger if a measurement of neutron
star rotation period is also available \citep[][and Figure \ref{fig:cdiagram} this work]{Corbet1986}.
The pulse periods of wind-accretion supergiant systems are generally
rather long ($\gtrsim$ 100 s){\mybf ,} with relatively short orbital periods ($\lesssim$ 20 days).
Roche-lobe overflow powered HMXBs are highly luminous,
with rather short pulse periods. Be star systems have longer
orbital periods ($\gtrsim$ 20 days), and exhibit a range of pulse periods that, while
correlated with orbital period \citep{Corbet1984,Corbet1986}, exhibit considerable
scatter around the correlation line. The Be system SAX J2103.5+4545, in particular,
is located far from the general correlation trend with a short orbital period, but a relatively
long spin period, more typical of a wind accretion supergiant system \citep[e.g.][and references therein]{Reig2010}.
For Be star systems,
there also appears to be a bimodal distribution of spin and orbital
periods \citep{Knigge2011} which has been attributed to
the neutron stars being formed in two types of supernova
\citep{Knigge2011} or differences in accretion modes \citep{Cheng2014}.

\begin{figure}
\epsscale{0.8}
\includegraphics[width=7.25cm,angle=0]{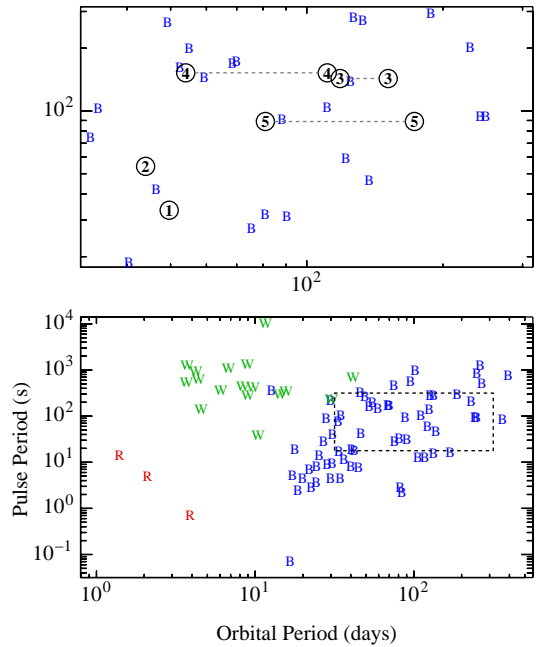}
\caption{Bottom panel: Pulse period vs. orbital period
for high-mass X-ray binaries. ``B'' = Be star
systems, ``W'' = sources accreting from a stellar
wind, ``R'' = sources thought to contain primary stars
at least close to filling their Roche lobes.
Top panel: Region of the pulse period vs. orbital period diagram covering
the sources discussed in this paper. The plot area corresponds to the dashed
box in the bottom panel. The known or possible parameters
are indicated with the circled numbers:
1 - \igr1448; \Porb\ = 49.64 d, \Ps\ = 33.4 s (this work);
2 - \axj1700; \Porb\ = 44.02 d, \Ps\ = 54 s \citep{Markwardt2010};
3 - Swift J1816.7-1613; \Porb\ = 118.5 days \citep{LaParola2014} or 151.1 days (this work), \Ps\ = 143 s \citep{Halpern2008,Krimm2013};
4 - AX J1820.5-1434; \Porb\ = 111 days (this work) or 54 days \citep{Segreto2013}, \Ps\ = 152.26 s   \citep{Kinugasa1998};
5 - XTE J1906+09; \Porb\ = 81 or 173 days (this work), \Ps\ = 89 s \citep{Wilson2002}.
}
\label{fig:cdiagram}
\end{figure}

Analysis of the BAT light curves for {\newbf \igrj1448} and
AX J1700.2-4220 shows strong persistent modulation at periods
of 49.6 and 44 days, respectively,
which are naturally
interpreted as the orbital periods of these systems.
The lengths of the orbital periods suggest that the modulations are ``Type I''
outbursts from Be star systems that occur near periastron passage.
{\newbf For \igrj1448, the possible 33.4 s pulse period in the \Swift\ XRT data would be
consistent with the Be star pulse/orbital period correlation
(Figure \ref{fig:cdiagram}) for a 49.6 day orbital period.
However, the significance of the modulation is very low, and is found
for the fainter source in the region which previous authors considered
to be the less likely counterpart \citep{Landi2009,Rodriguez2010}.
Additional
observations are therefore required to determine the reality of the candidate pulse period,
and to search for shorter periods for both XRT sources in the region 
than was possible with the 2.5s time resolution.}
For AX J1700.2-4220, the 54 s pulse period found by \citet{Markwardt2010},
combined with the 44 day period again clearly place the source in the
Be star region of the pulse/orbital period diagram (Figure \ref{fig:cdiagram}).

For \Swift\ J1816.7-1613, AX J1820.5-1434, and XTE J1906+090
the situation is more complicated.
All three of these sources are pulsars, and the expectation
is that they are probably Be star sources.
While all three show at least one prominent peak in their power spectra,
there can be problems in interpreting these as the orbital periods of these 
systems.

For Swift J1816.7-1613, the power spectrum of the BAT light curve shows
a modest significance peak near 151 days. However, this is not consistent
with the results of \citet{LaParola2014} who instead find a 118.5 $\pm$ 0.8 d
period, also from BAT data. The analysis of \citet{LaParola2014} differs from
ours in that they use a folding analysis, and also only selected 200 day long
sections of the light curve centered on apparent outbursts in the light curve.
Our analysis of the same time ranges used by \citet{LaParola2014}
using a power spectrum also shows its strongest peak near 118.5 days.
This indicates that the difference in our results is primarily due to the use
by \citet{LaParola2014} of only restricted times while we use the full light curve.
We regard the orbital period of Swift J1816.7-1613 as not yet clearly determined.
It is possible that neither the 151 day nor the 118.5 day modulations are the
actual orbital period, and both may be due to non-periodic variability.

For AX\,J1820.5-1434, it may be possible to interpret the 111 day period
as the orbital period, and the previously proposed 54 day period \citep{Segreto2013} as being
due to the detection of the harmonic of this period.
However, the overall modulation seen in the folded light curve 
is not smooth, which may hint at more complicated behavior and
possibly that the flux modulation period is not the actual orbital
period of the system. Indeed, the time-sliced folded light curves (Figure \ref{fig:1820_slice_fold}{\mybf )}
are suggestive of a change in modulation period 
from \sqig 54 days to \sqig 111 days {\viibf after \sqig MJD 55,200}.
{\viibf Such a change in period might be analogous to that
exhibited by the peculiar HMXB 4U\,2206+54, where an initial apparent
period of 9.6 days was later found to transition to modulation at
double this period \citep{Corbet2007a,Wang2009,Levine2011}.}

In the case of XTE J1906+090, where two peaks are seen in the power spectrum,
it is unclear which, if either, is the orbital period.
The apparent conflict with the pulse-timing results, if one of these peaks
is the orbital period, is not necessarily a problem. As noted by
\citet{Wilson2002}, the apparent orbital modulation of the pulse frequency
was based on a single 25 day interval which would require additional
data to confirm their parameters. 

For Swift J1816.7-1613, AX J1820.5-1434, and XTE J1906+090, pulse timing might
have the potential to allow an unambiguous determination of these sources'
orbital periods \citep[see e.g.][]{Townsend2011}.
In addition, if definite optical counterparts can be determined, then
long-term optical photometry might {\mybf also} enable detection of their orbital periods
\citep[see e.g.][and references therein]{Charles2012}.
{\mybf Further}, with the determination of definite optical counterparts, long-term
studies of \halpha\ could be undertaken which would be a diagnostic of the Be
star's decretion disk. Changes in X-ray modulation could then be investigated
for connections with decretion disk properties. For example, changes in
disk size for AX J1820.5-1434 could account for the factor \sqig2 change
in modulation period{\newbf ,} if the size changes result in the neutron star
impacting the decretion disk either once or twice per orbit.

\newpage
\section{Conclusion}

With over 11 years of all-sky observations at X-ray energies
above 15 keV, the \Swift\ BAT light curves provide a sensitive
way of investigating the long-term variability of HMXBs.
The five sources considered here demonstrate the wide diversity
of different types of long-period modulation that can be exhibited
by HMXBs.
In two cases {\mybf ({\newbf \igrj1448} and
AX J1700.2-4220),} orbital periods can clearly be determined,
which are important to understanding the properties of these
systems. In {\mybf the three} other {\mybf systems}, while periodic or quasi-periodic
long-term variation is seen, its interpretation is less clear. {\mybf It is possible that
none of these systems are exhibiting orbital modulation of their X-ray fluxes,
and that more complicated variability is occurring.}
For these systems, the continued accumulation of long-term
light curves and other types of observations, including at
other wavelengths, will be important to determining the
physical driving mechanism(s) behind the variability.

\acknowledgements

We thank Colleen Wilson-Hodge for useful discussions on
XTE J1906+090. We thank two anonymous referees for helpful comments.
This work was supported by NASA grant 14-ADAP14-0167.
JBC was supported by an appointment to the NASA Postdoctoral Program at the
Goddard Space Flight Center administered by Universities Space Research Association
through a contract with NASA.

\pagebreak


\end{document}